\documentclass[]{book}
\usepackage{graphicx,subfigure}
\usepackage{dcolumn}
\usepackage{bm}
\usepackage{amsmath}
\usepackage{amssymb}
\usepackage{chapterbib}
\usepackage{color}

\begin{document}

\chapter{Critical brain dynamics at large scale}
{\flushleft
{Dante R. Chialvo}

{CONICET, (Consejo Nacional de Investigaciones Cient\'{i}ficas y Tecnol\'{o}gicas), Buenos Aires, Argentina.}}
\vspace*{5mm}
\begin{quote}{\it Essentially, all modeling of brain function from studying models of neural networks has ignored the self-organized aspects of the process, but has concentrated on designing a working brain by engineering all the connections of inputs and outputs.}{Per Bak\cite{Bak}\\}
\end{quote}
\vspace*{5mm}
\small{{\it Abstract} -
Highly correlated brain dynamics produces synchronized states with no behavioral value, while weakly correlated
dynamics prevent information flow.  In between these states, the unique dynamical features of the critical state endow the brain with properties which are fundamental for adaptive behavior. We discuss the idea put forward two decades ago by Per Bak that the working brain stays at an intermediate (critical) regime characterized by power-law correlations. This proposal is now supported by a wide body of empirical evidence at different scales demonstrating that the spatiotemporal brain dynamics exhibit key signatures of critical dynamics, previously recognized in other complex systems. The rationale behind this program is discussed in these notes, followed by an account of the most recent results. \let\thefootnote\relax\footnote{In ``Criticality in Neural Systems'', Niebur E, Plenz D, Schuster HG. (eds.) 2013 (in press).}
}
 
\section{Introduction: If criticality is the solution, what is the problem?}

Criticality, in simple terms, refers to a distinctive set of properties found only at the boundary separating regimes with different dynamics, for instance between an ordered and a disordered phase. The dynamics of critical phenomena are a peculiar mix of order and disorder, whose detailed understanding constitute one of the mayor achievements of statistical physics in the last century \cite{stanley}.
  
What is the problem for which critical phenomena can be relevant in the context of the brain? The first problem is to understand how the \emph{very large} conglomerate of interconnected neurons produce a \emph{wide repertoire} of behaviors in a \emph{flexible} and self organized way. This issue is not resolved at any rate, demonstrable by the fact that detailed models constructed to account for such dynamics fail at some of the three \emph{emphasized} aspects: Either 1) the model is an unrealistic low dimensional version of the neural structure of interest;  or 2) it produces a single behavior (i.e., a hardwired circuit); and consequently 3) it cannot flexibly perform more than one simple thing. A careful analysis of the literature will reveal that only by arbitrarily changing the neuronal connections, current mathematical models can play a reasonable wide repertoire of behaviors. Of course, this rewiring implies a kind of supplementary brain governing which connections need to be rewired in each case. Consequently,  generating behavioral variability out of the same neural structure is a fundamental question  which is screaming to be answered,  but seldom is even being asked.
 
A  second related problem is how stability is achieved in such a very large system with astronomical number of neurons, each one continuously receiving thousands of inputs from other neurons. We still lack a precise knowledge of how the cortex prevents an explosive propagation of activity while still managing to share information across areas. It is obvious that if the average number of neurons activated by one neuron is too high (i.e., supercritical)  a massive activation of the entire network will ensue, while if it is too low (i.e., subcritical), propagation will die out.  
It was Turing, about fifty years ago \cite{turing}, the first to speculate that the brain, in order to work properly, needs to be at  a critical regime, i.e., one in which these opposing forces are balanced. 

Criticality as a potential solution to these issues  was first explored by Per Bak \cite{Bak} and colleagues \cite{Stepha,ceccatto,Chialvo99,Bak2001,wakelingbak01}
while attempting to apply ideas of self-organized criticality \cite{Bak87,Jensen} to the study of living systems.  Throughout the last decade of his short but productive life, in uncountable lively lectures,  Bak enthusiastically broadcasted the idea  that if the world at large is studied as any other complex system, it will reveal a variety of instances in which critical dynamics will be recognized as the relevant phenomena at play. Basically, the emphasis was  in considering criticality as another attractor. The claim was that {\it``dynamical systems with extended spatial degrees of freedom naturally evolve into self-organized critical structures of states which are barely stable. The combination of dynamical minimal stability and spatial scaling leads to a power law for temporal fluctuations''} \cite{Bak87}. 

These ideas were only a portion of Bak's much broader and deeper insight about how Nature works in general, often communicated in his unforgiving way, as for instance  when challenging colleagues by asking: {\it ``Is biology too difficult for biologists?  And what can physics, dealing with the simple and lawful, contribute to biology, which deals with the complex and diverse. These complex many-body problems might have similarities to problems studied in particle and solid-state physics."} \cite{bak1}. Thus, Per Bak was convinced that the critical state was a novel dynamical attractor to which large distributed systems will eventually converge, given some relatively simple conditions. From this viewpoint, the understanding of the brain belongs to the same problem of understanding complexity in Nature. 

The above comments should inspire us to think again about the much larger question underlying the study of  brain dynamics using ideas from critical phenomena. Bak's (and colleagues) legacy will be incomplete if we restrict ourselves (for instance) to find power laws in the brain and  compare it in health and disease. By its theoretical foundations, critical phenomena offers the opportunity to understand how the brain works, in the same magnitude that it impacted in some other areas, as for instance in the mathematical modeling of Sepkoski fossil record of species extinction events, which opened a completely novel strategy to study how macroevolution works \cite{baksneppen}. 

The remaining of these notes are dedicated to review recent work on large scale brain dynamics inspired on Bak's ideas. The material is organized as follows:  the next section dwells into what is essentially novel about critical dynamics; Sections 3 and 4 are dedicated to discuss how to recognize criticality. Section 5  discusses the main implications of the results presented and Section 6 close the chapter with a summary.
 
\section{What is criticality good for?} 
 
According to this program the methods used in physics to study the properties of matter should be useful to characterize brain function \cite{chialvo2010}. How reasonable is that?  A simple but strong assumption needs to be made: that the mind is nothing more than the emergent global dynamics of neuronal interactions, in the same sense than ferromagnetism is an emergent property of the interaction between neighboring spins  and an external field. To appreciate the validity of this point a key result from statistical physics is relevant  here: universality. In brief, this notion says that a huge family of systems will follow the same laws and exhibit the same dynamics providing that some set of minimum conditions are meet. This conditions involve only the presence of some nonlinearity, under some boundary conditions and some type of interactions. Any other details of the system will not be relevant, meaning that the process will arise in the same quanti- and qualitative manner in very diverse systems, where order, disorder or the observation of one type of dynamics over another will be dictated by the  strength and type of the interactions. This is seen throughout nature, from  cell function (warranted by the interaction of multiple metabolic reactions) to global macroeconomics (modulated by trade), and so on. 

Perhaps, considering the unthinkable one could appreciate better what universality means, in general, and later translate it to complex systems. The world would be a completely different place without universality, imagine if each phenomena would be explained by a different ``relation'' (since it would not be possible to talk in terms of general laws) between intervening particles and forces. Gravity would be different for each metals or different materials, Galileo's experiments would not repeat themselves unless for the same material he used, etc. It can be said that without universality, each phenomena we are familiar with would be foreign and strange.

\subsection{Emergence}
Throughout nature, it is common to observe similar collective properties emerging independently of the details of each system. But what is emergence and why is relevant to discuss it in this context?  
Emergence refers to the unexpected collective spatiotemporal patterns exhibited by \emph{large} complex systems. In this context, ``unexpected'' refers to our inability (mathematical and otherwise) to derive such emergent patterns from the equations describing the dynamics of the individual parts of the system. As discussed at length elsewhere \cite{Bak, Bak95}, complex systems are usually \emph{large} conglomerates of \emph{interacting} elements, each one exhibiting some sort of \emph{nonlinear} dynamics. Without entering into details, it is also known that the interaction can also be indirect, for instance through some mean field. Usually energy enters into the system, thus some sort of driving is present. The three \emph{emphasized} features, ( i.e., large number of interacting nonlinear elements)  are necessary, although not sufficient, conditions for a system to exhibit emergent complex behavior at some point.

As long as the dynamics of each individual element is nonlinear, other details of the origin  and nature of the nonlinearities are not important \cite{Bak,anderson}. For instance, elements can be humans, driven by food and other energy resources, from which some collective political or social structure eventually arises. It is well known that,  whatever the type of structure that emerges, it is unlikely to appear if one of the three above-emphasized properties is absent. Conversely, the interaction of a {\it small number} of  {\it linear} elements won't produce any of this ``unexpected'' complex behavior (indeed this is the case in which everything can be mathematically anticipated). 

\subsection{Spontaneous brain activity is complex}
It is evident, from the very early electrical recordings a century ago, that the brain is spontaneously active, even in absence of external inputs. However obvious this observation could appear, it was only recently that the dynamical features of the spontaneous brain state started to be studied in any significant way. 

Work on brain rhythms at small and large brain scales shows that spontaneous healthy brain dynamics is not composed by completely random activity patterns nor by periodic oscillations \cite{Buzsaki}. Careful analysis of the statistical properties of neural dynamics under no explicit input has identified complex patterns of activity previously neglected as background noise dynamics. The fact is that brain activity is always essentially arrhythmic regardless of how it is monitored, whether as electrical activity in the scalp (EEG), by techniques of functional magnetic resonance imaging (fMRI), in the synchronization of oscillatory activity \cite{Linkenkaer,Stam}, or in the statistical features of local field potentials peaks \cite{Plenz2007}.

It has been pointed out repeatedly \cite{Bullock,Logothetis, Eckhorn, Miller, Manning} that, under healthy conditions, no brain temporal scale takes primacy over average, resulting in power spectral densities decaying of ``1/f noise''. Behavior, the ultimate interface between brain dynamics and the environment, also exhibits scale invariant features as shown in human cognition \cite{Gilden,Maylor,Ward} human motion \cite{Nakamura} as well as animal motion \cite{Anteneodo}. The origin of the brain scale free dynamics was not adequately investigated until recently, probably (and paradoxically) due to the ubiquity of scale invariance in nature \cite{Bak}. The potential significance of a renewed interpretation of the brain spontaneous patterns in term of scale invariance is at least double. On one side,  it provides important clues about brain organization, in the sense that our previous ideas cannot easily accommodate these new findings. Also, the class of complex dynamics observed seems to provide the brain with previously unrecognized robust properties. 
 
\subsection{Emergent complexity is always critical}
The commonality of scale-free dynamics in the brain naturally leads one to ask what physics knows about very general mechanisms able to produce such dynamics. Attempts to explain and generate nature's non- uniformity included several mathematical models and recipes, but few succeeded in creating complexity without embedding the equations with complexity. The important point is that  including the complexity in the model will only result in a simulation of the real system, without entailing any understanding of complexity. The most significant efforts were those aimed at discovering the conditions in which something complex emerges from the interaction of the constituting non-complex elements \cite{Bak, Bak87}. Initial inspiration was drawn from work in the field of phase transitions and critical phenomena. Precisely, one of the novelties of critical phenomena is the fact that out of the short-range interaction of simple elements eventually long-range spatiotemporal correlated patterns emerge. As such, critical dynamics have been documented in species evolution \cite{Bak}, ants collective foraging \cite{Beckers,Beekman} and swarm models \cite{Rauch}, bacterial populations \cite{Nicolis}, traffic flow in highways \cite{Bak} and on the Internet  \cite{Takayasu}, macroeconomic dynamics \cite{Lux}, forest fires \cite{Malamud}, rainfall dynamics \cite{Peters1, Peters2, Peters3} and flock formation \cite{Cavagna}. Same rationale leads to the conjecture \cite{Bak, Chialvo99,Bak2001} that also the complexity of brain dynamics is just another signature of an underlying critical process. Since at the point near the transition the largest number of metastable states exists, the brain can then access the largest repertoire of behaviors in a flexible way. That view claimed that the most fundamental properties of the brain only are possible staying close to that critical instability  independently of how such state is reached or maintained. In the following sections recent empirical evidence supporting this hypothesis will be discussed. 

\begin{figure}[h]
\centering
\includegraphics[angle=0,width=0.60\textwidth,keepaspectratio,clip]{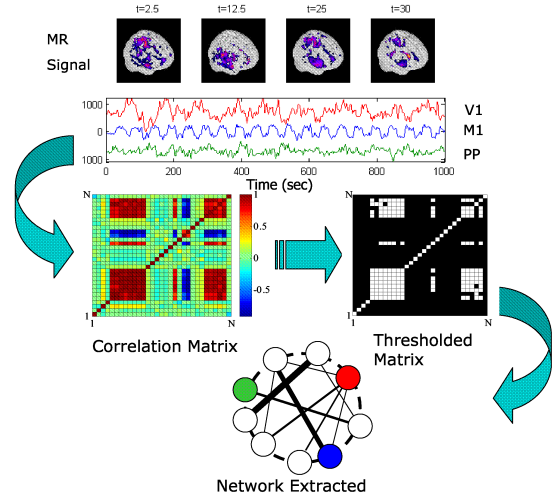}
\caption{ Methodology used to extract functional networks from the brain fMRI  BOLD signals.
The correlation matrix is calculated from all pairs of BOLD time series. The strongest correlations are selected to define the networks nodes. Top four images represent examples of snapshots of activity at one moment and the three traces correspond to time series of activity at selected voxels from visual (V1), motor (M1) and posterio-parietal
(PP) cortices. Figure redrawn from \cite{eguiluz}
\label{figeguiluz}}
\end{figure}
\section{Statistical signatures of critical dynamics} 

The presence of scaling and correlations spanning the size of the system are usually hints of critical phenomena. While, in principle, it is relatively simple to identify these signatures, in the case of finite data and the absence of a formal theory, as is the case of the brain, any initial indication of criticality need to be checked against many known artifacts.
In the next paragraphs we discuss the most relevant efforts to identify these signatures in large scale brain data.
 
\subsection{Hunting for power laws in densities functions}

The dynamical skeleton of a complex system can be derived from its correlation network, i.e., the subsets of the nodes linked by some minimum correlation value (computed from the system activity). 
As early as 2003 Eguiluz and colleagues \cite{eguiluz} used functional magnetic resonance imaging (fMRI) data to extract for the very first {\em functional networks} connecting correlated human brain sites. Networks were constructed (see Fig. \ref{figeguiluz}) by connecting the brain sites with strongest correlations between their blood oxygenated level dependent (BOLD) signal. 
The analysis of the resulting networks in different tasks showed that: (a) the distribution of functional connections, and the probability of finding a link vs. distance were both scale-free, (b) the characteristic path length was small and comparable with those of equivalent random networks, and (c) the clustering coefficient was orders of magnitude larger than those of equivalent random networks. It was suggested that these properties, typical of scale-free small world networks, should reflect important functional information about brain states and provide mechanistic clues.

\begin{figure}[h]
\begin{center}
\includegraphics[angle=0,width=0.60\textwidth,keepaspectratio,clip]{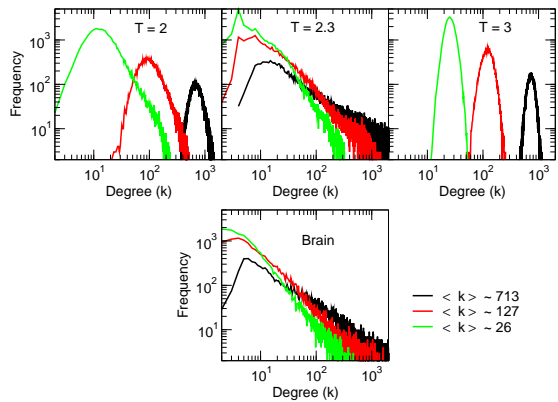}
\end{center}
\caption{At criticality, brain and Ising networks are indistinguishable from each other. The graphs show a comparison of the link density distributions computed from correlation networks extracted from brain data (bottom panel) and from  numerical simulations of the Ising model (top three panels) at three temperatures: critical ($T=2.3$), sub ($T=2$) and supercritical ($T=3$). Top three panels depict the degree distribution for the Ising
networks at $T=2$, $T=2.3$ and $T=3$ for three representative
values of $\langle k \rangle \approx 26$, $127$, and $ 713$.
Bottom panel: Degree distribution for correlated brain
network for the same three values of $\langle k \rangle$. 
Figure redrawn from Fraiman et al \cite{Fraiman2008}.
\label{figfraimanising}}
\end{figure}

This was investigated in a subsequent paper by Fraiman et al. \cite{Fraiman2008} who studied the  dynamic of the spontaneous (i.e., at ``rest'') fluctuations of brain activity with fMRI.  Brain ``rest'' is defined -more or less unsuccessfully- as the state in which there is no explicit brain input or output. Now is widely accepted that the structure and location of large-scale brain networks can be derived from the interaction of cortical regions during rest which closely match the same regions responding to a wide variety of different activation conditions \cite{fox2007,Beckmann-2009}. These so-called resting state networks (RSN) can be reliably computed from the fluctuations of the BOLD signals of the resting brain, with great consistency across subjects \cite{beckmann2005,xiong,cordes} even during sleep \cite{fuku} or anesthesia \cite{vincent}.
Fraiman et al \cite{Fraiman2008} focused on the question of whether such states can be comparable to any known \emph{dynamical} state. For that purpose, correlation networks from human brain fMRI were contrasted with correlation networks extracted from numerical simulations of the Ising model in 2D, at different temperatures. For the critical temperature $T_c$, striking similarities (as shown in Fig. \ref{figfraimanising}) appear in the most relevant statistical properties, making the two networks indistinguishable from each other. These results were interpreted as lending additional support to the conjecture that the dynamics of the functioning brain is near a critical point.

Kitzbichler et al \cite{Kitz} analyzed functional MRI and MEG data recorded from normal volunteers at resting state using phase synchronization between diverse spatial locations. They reported a scale invariant distribution for the length of time that two brain locations on the average remained locked. This distribution was also found in the Ising and the Kuramoto model \cite{Kuramoto} at the critical state, suggesting that the data exhibited criticality.
This work was revisited recently by Botcharova et al. \cite{botcharova} who investigated whether the display of  power law statistics of the two measures of synchronization - phase locking intervals and global lability of synchronization -  can be analogous to similar scaling at the critical threshold in classical models of synchronization. Results confirmed only partially the previous findings, emphasizing the need to proceed with caution in making direct analogies between the brain dynamics and systems at criticality.   Specifically, they showed that  ``the pooling of pairwise phase-locking intervals from a non-critically interacting system can produce a distribution that is similarly assessed as being power law. In contrast, the global lability of synchronization measure is shown to better discriminate critical from non critical interaction'' \cite{botcharova}.

\begin{figure}[tb]
\center
\includegraphics[width=0.50\textwidth,keepaspectratio,clip]{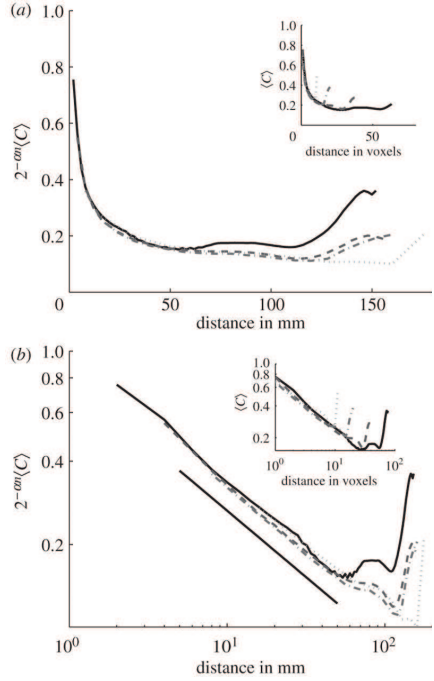}
\caption{Self-similarity of the brain fMRI two-correlation function.  
The plot shows the renormalized average correlation function versus distance for the four levels of description: 
solid line: 128 x 128 x31 (n=0); dashed line: 64x64x16(n=1); dahsed-dotted, 32 x 32 x 8 (n=2); and dotted line: 16x16x4 (n= 3). (a) Linear-linear and (b) log-log axis.  The exponent $\beta= 0.47 +/- 0.2$ describes well the data. Figure redrawn from Expert et al. \cite{Expert}
\label{figexpert}}
\end{figure}

The works commented up until now rely on determining if probability density functions (i.e., node degree, or synchronization lengths) obey power laws. The approach from Expert et al. \cite{Expert} looked at a well known property of the dynamics at criticality: self-similarity.  They investigated whether the two point correlation function can be renormalized.  This is  a very well understood technique used in critical phenomena in which the data sets are coarse grained at successive scales while computing some statistic. They were able to show that the two point correlation function of the BOLD signal is invariant under changes in the spatial scale as shown in Fig. \ref{figexpert}, which together with the temporal $1/f$ scaling exhibited by BOLD time series, suggests critical dynamics. 
 
\subsection{Beyond fitting: variance and correlation scaling of brain noise}
An unexpected new angle into the problem of criticality was offered by the surging interest in the source of  the BOLD signal variability and its information content. For instance, it was shown  recently \cite{noise} in a group of subjects of different age, that the BOLD signal standard deviation can be a better predictor of the subject age than the average.  Furthermore, additional work focused on the relation between the fMRI signal variability and a task performance, concluded that faster and more consistent performers exhibit significantly higher brain variability across tasks than the poorer performing subjects \cite{Garrett}. Overall, these  results suggested that the understanding of the brain resting dynamics can benefit from a detailed study of the BOLD variability {\it per se}.  

Precisely at this aim was directed the work in \cite{Fraiman2012},  which studied the statistical properties of the spontaneous BOLD fluctuations and its possible dynamical mechanisms. In these studies, an ensemble of brain regions of different sizes  were defined and the statistics of the fluctuations and correlations were computed  as a function of the region's size. The report identifies anomalous scaling of the variance as a function of the number of elements and a distinctive divergence of the correlations with the size of the cluster considered. We now proceed to describe these findings in detail.
\subsubsection*{Anomalous scaling:}
The object of interest are the fluctuations of the BOLD signal around its mean, which for the thirty-five RSN clusters used by \cite{Fraiman2012}, are defined as
\begin{equation}
{B_h}(\vec{x}_i,t)=
B(\vec{x}_i,t)-\frac{1}{N_H}\overset{N_H}{\underset{i=1}{\sum}} B(\vec{x}_i,t),
\end{equation}
where $\vec{x}_i$ represents the position of the voxel $i$ that belongs to the cluster $H$ of size $N_H$. 
These signals will be used to study the correlation properties of the activity in each cluster.

The mean activity of each $h$ cluster is defined as
\begin{equation}
\overline{B}(t)= \frac{1}{N_H}\overset{N_H}{\underset{i=1}{\sum}} B(\vec{x}_i,t),
\end{equation}

and its variance is defined as 
\begin{equation}
\sigma_{_{\overline{B}(t)}}^2=\frac{1}{T}\overset{T}{\underset{t=1}{\sum}}(\overline{B}(t)-\overline{\overline{B}})^2,
\end{equation}
where $\overline{\overline{B}}=\frac{1}{T}\overset{T}{\underset{t=1}{\sum}}\overline{B}(t)$ and $T$ the number of temporal points. Please notice that the average subtracted  in Eq. 1 is the mean  at time $t$  (computed over $N$ voxels) of the BOLD signals, not to be confused with the BOLD signal averaged over $T$ temporal points. 

\begin{figure}[tp]
\center
\includegraphics[width=0.95\textwidth,keepaspectratio,clip]{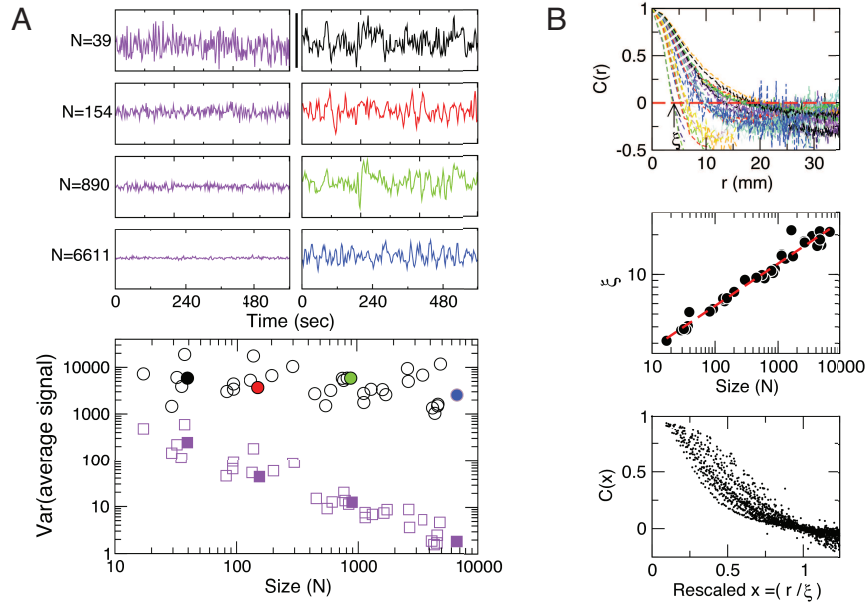}
\caption{Spontaneous fluctuations of fMRI data shows anomalous behavior of the variance (A) and divergence of the correlation length (B).  
Top figures in Panel A show four examples of average BOLD time series (i.e., $\overline{B}(t)$  in Eq. 2) computed from clusters  of different sizes $N$. Note that while the amplitude of the raw BOLD signals (right panels) remains approximately constant, in the case of the shuffled data sets (left panels) the amplitude decreases drastically for increasing cluster sizes.
The bottom graph in  Panel A shows the calculations for the thirty five clusters (circles) plotted as a function of the cluster size demonstrating that variance is independent of the RSN's cluster size. The squares symbols show similar computations for a surrogate time series constructed by randomly reordering the original BOLD time series, which exhibit the expected $1/N$ scaling (dashed line). Filled symbols in bottom panel are used to denote the values for the time series used as examples in the top panel.
In panel B there are three graphs: the top one shows the correlation function $C(r)$ as a function of  distance for clusters of different sizes. Contrary to naive expectations, large clusters are as correlated as relatively smaller ones: the correlation length increases with cluster size, a well known signature of criticality.  Each line in the top panel shows the mean cross-correlation $C(r)$ of BOLD activity fluctuations as a function of distance $r$ averaged over all time series of each of the thirty five clusters. The correlation length $\xi$, denoted by the zero crossing of $C(r)$ is not a constant. As shown in the  middle graph  scale  $\xi$ grows linearly with the average cluster' diameter $d$ for all the thirty five clusters (filled circles), $\xi \sim N^{1/3}$. The bottom graph shows the collapse of $C(r)$ by rescaling the distance with $\xi$. Figure redrawn from Fraiman et al \cite{Fraiman2012}
\label{figfraiman2012A}}
\end{figure}

Since the BOLD signal fluctuates widely and the number $N$ of voxels in the clusters can be very large, one might expect that the aggregate of Eq. 1 obeys the law of the large numbers. If this was true, the variance of the mean field $ \sigma_{_{\overline{B}(t)}}^2$ in Eq. 3 would decrease with $N$ as $N^{-1}$. 
In other words one  would expect a smaller amplitude fluctuation for the average BOLD signal recorded in clusters (i.e., $\overline{B}(t)$)   comprised by large number of voxels compared with smaller clusters. 
However, the data in Fig. \ref{figfraiman2012A}A  shows otherwise, the variance of the average activity remains approximately constant over a change of four orders of magnitude in cluster' sizes. The strong departure from the $N^{-1}$ decay is enough to disregard further statistical testing. which is confirmed by recomputing  the variance for artificially constructed clusters having similar number of voxels but composed of the randomly reordered $B_k(t)$ BOLD raw time series (as the four examples in the top left panels of Fig. \ref{figfraiman2012A}A). As expected, in this case the variance  (plotted using squares symbols in the bottom panel of Fig. \ref{figfraiman2012A}A ) obeys the $N^{-1}$ law). 

\subsubsection*{Correlation length:} A straightforward  approach to understand the correlation behavior commonly used in large collective systems \cite{Cavagna} is to determine the correlation length at various system's sizes. The correlation length is the average distance at which the correlations of the fluctuations around the mean crosses zero. It describes how far one has to move to observe any two points in a system behaving independently of each other. Notice that, by definition, the computation of the correlation length is done over the fluctuations around the mean, and not over the raw BOLD signals, otherwise global correlations may produce a single spurious correlation length value commensurate with the brain size. 

Thus, we start by computing for each voxel BOLD time series their fluctuations around the mean of the cluster that they belong. Recall the expression in Eq. 1.1,
where $B$ is the BOLD time series at a given voxel and $\vec{x}_i$ represents the position of the voxel $i$ that belongs to the cluster $H$ of size $N_H$. By definition
the mean of the BOLD fluctuations of each cluster vanishes,
\begin{equation}
\overset{N_k}{\underset{i=1}{\sum}} \overline{B_h}(\vec{x}_i,t)=0 \quad \quad \forall t.
 \end{equation}
Next we compute the average correlation function of the BOLD fluctuations
 between all pairs of voxels in the cluster considered, which are separated by a distance $r$:
{\tiny
\begin{equation}
\langle C_H(r)\rangle=<\frac{(B_H(\overrightarrow{x},t)-<B_h(\overrightarrow{x},t))>_t)(B_H(\overrightarrow{x}+r\overrightarrow{u},t)-<B_h(\overrightarrow{x}+r\overrightarrow{u},t)>_t)}{(<B_H(\overrightarrow{x},t)^2>_t-<B_H(\overrightarrow{x},t)>_t^2)^{1/2}(<B_H(\overrightarrow{x}+r\overrightarrow{u},t)^2>_t-<B_H(\overrightarrow{x}+r\overrightarrow{u},t)>_t^2)^{1/2}}>_{t,\overrightarrow{x},\overrightarrow{u}}
\end{equation}
}
where $\vec{u}$ is a unitary vector, and $\langle . \rangle_{w}$ represent averages over $w$.
  
The typical form we observe for $C(r)$  is shown in the top panel of Fig. \ref{figfraiman2012A} B. The first striking feature to note is the absence of a unique $C(r)$ for all clusters. Nevertheless, they are qualitatively similar, being at short distances close to unity, to decay as $r$ increases, and then becoming negative for longer voxel-to-voxel distances. Such behavior indicates that within each and any cluster, on the average, the fluctuations around the mean are strongly positive at short distance and strongly anti-correlated at larger distances, whereas there is no range of distance for which the correlation vanishes.

It is necessary to clarify whether the $\xi$ divergence is trivially determined by the structural connectivity. In that case $C$ must be constant throughout the entire recordings. Conversely, if the dynamics are critical, their average value will not be constant, since it is the product of a combination of some instances of high spatial coordination intermixed with moments of dis-coordination.  
In order to answer this question we study the mean correlation $\langle C \rangle$ as a function of time for regions of interest of various sizes, for non-overlapping periods of 10 temporal points.
\begin{figure}[ht]
\centerline{
\includegraphics[width=.7\textwidth,clip=true]{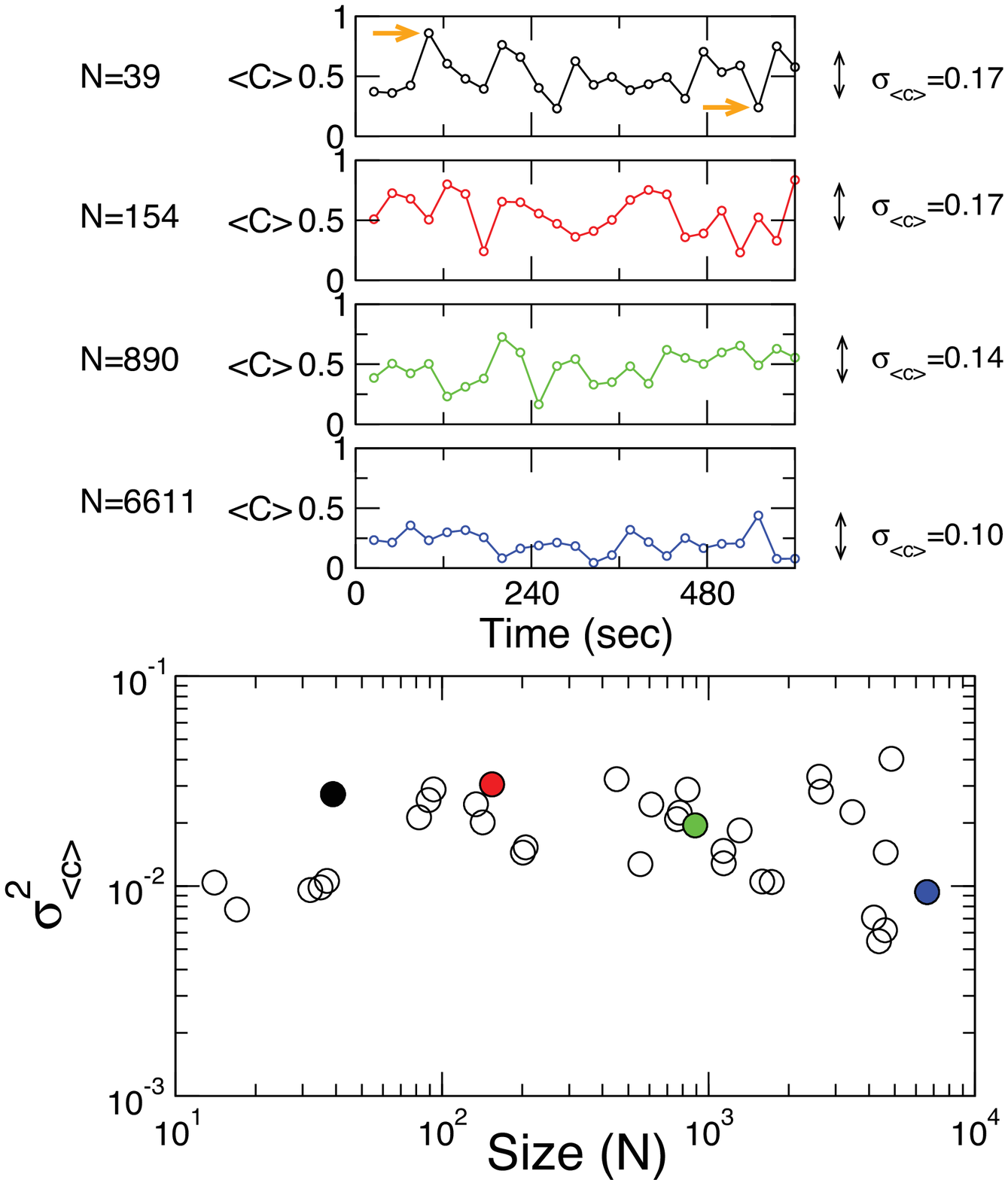}}
 \caption{Bursts of high correlations are observed at all cluster sizes, resulting in approximately the same variance, despite the four orders of magnitude change in the cluster size. The top panels illustrate representative examples of short-term mean correlation  $\langle C \rangle$ of the BOLD signals as a function of time for four sizes spanning four orders of magnitude.  The arrows show examples of two instances of highly correlated and weakly correlated activity, respectively. Bottom panel shows the variance of $\langle C \rangle$ as a function of cluster sizes. The four examples on the top traces are denoted  with filled circles in the bottom plot. Figure redrawn from Fraiman et al \cite{Fraiman2012}
 \label{figfraiman2012B}
 }
 \end{figure}
 
Figure \ref{figfraiman2012B} shows the behavior of $\langle C \rangle$ over time for four different cluster's sizes. Notice that, in all cases, there are instances of large correlation followed by moments of week coordination, as those indicated by the arrows in the uppermost panel.  We have verified that this behavior is not sensitive to the choice of the length of the window in which  $\langle C \rangle$  is computed.
These bursts keep the variance of the correlations almost constant (i.e., in this example, there is a minor decrease in variance (by a factor of 0.4) for a huge increase in size (by a factor of 170). This is observed for any of the cluster sizes as shown in the bottom panel of Fig. 4 where the variance of $\langle C \rangle$  is approximately constant, despite the four order of magnitude increase in sizes.
The results of these calculations imply that independently of the size of the cluster considered, there is always an instance in which a large percentage of voxels are highly coherent and another instance in which each voxels activity is relatively independent.

Thus, to summarize Fraiman et al. work \cite{Fraiman2012}, revealed three key statistical properties of the brain BOLD signal variability:
 \begin{itemize}

\item{the variance of the average BOLD fluctuations computed from ensembles of widely different sizes remains constant, (i.e., anomalous scaling);}

\item{the analysis of short-term correlations reveals bursts of  high coherence between arbitrarily far apart voxels indicating that the variance anomalous scaling has a dynamical (and not structural) origin; }

\item{the correlation length measured at different regions increases with region's size, as well as its mutual information.}
\end{itemize}

\begin{figure}[htb]
\centerline{
\includegraphics[width=.75\textwidth,clip=true]{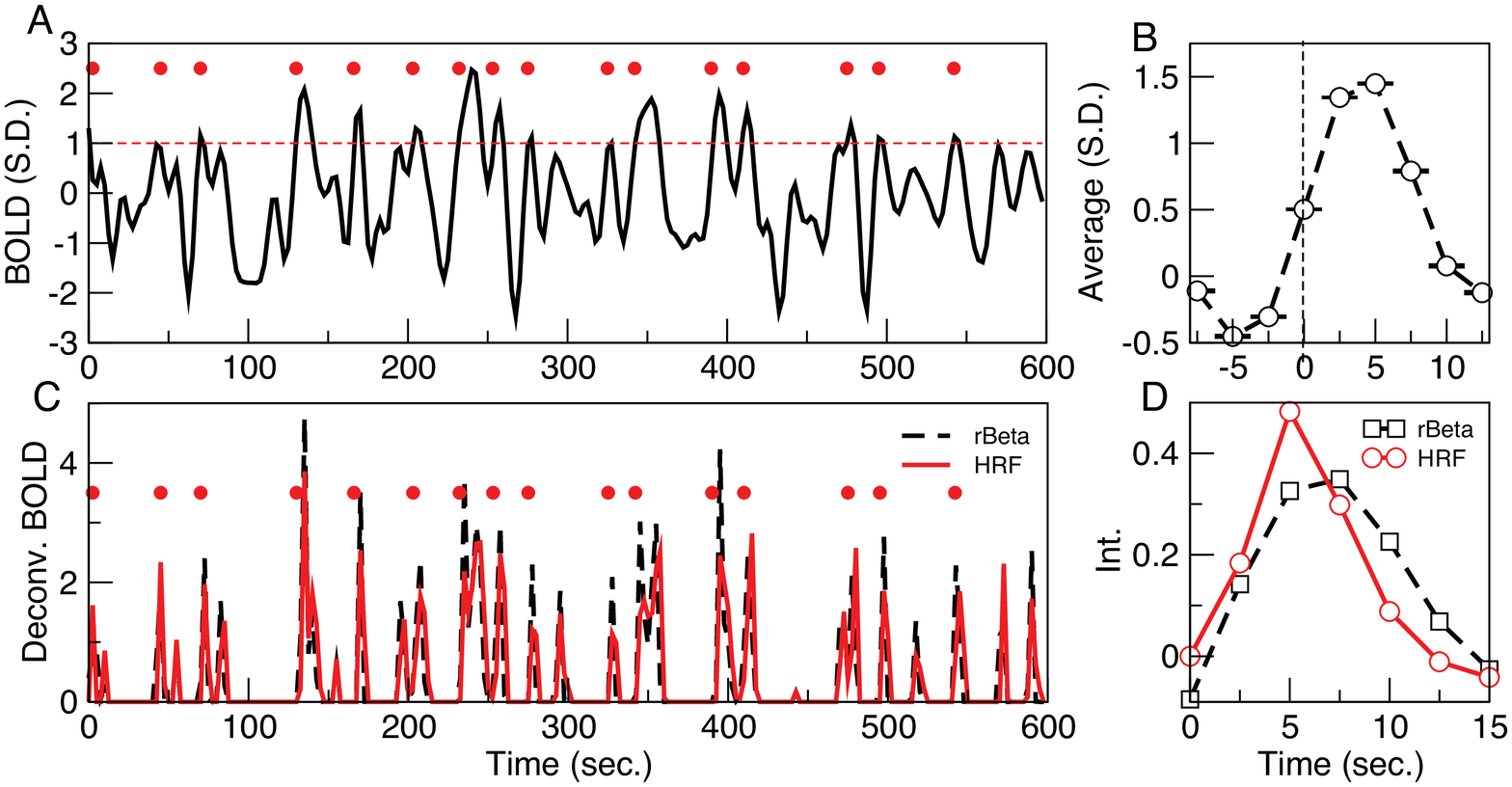}}
 \caption{ (A) Example of a point process (filled circles) extracted from the normalized BOLD signal. Each point corresponds to a threshold (dashed line at 1 S.D.) crossing from below. (B) Average BOLD signal (from all voxels of one subject) triggered at each threshold crossing. (C) The peaks of the de-convolved BOLD signal, using either the hemodynamic response function (HRF) or the rBeta function \cite{enzo2011} depicted in panel D,  coincide on a great majority with the timing of the points shown in panel A. Figure redrawn from Tagliazucchi et al \cite{enzo2012} 
 \label{figpuntos}}
 \end{figure}

\section{Beyond averages: Spatiotemporal brain dynamics at criticality}

Without exception, all the reports considering large scale brain critical dynamics resorted to the computation of averages over certain time and/or space scales. However, since time and space is essential for brain function,  it would be desirable to make statements of {\it where} and {\it when} the dynamics is at the brink of instability, i.e., the hallmark of criticality. In this section we summarize novel ideas that  attempt to meet this challenge by developing techniques that consider large-scale dynamics in space and time in the same way that climate patterns are dealt with, tempting us to call these efforts  ``brain meteorology''.

Tagliazucchi et al. departed from the current brain imaging techniques based on the analysis of \emph{gradual and continuous} changes in the  brain blood oxygenated level dependent (BOLD) signal. By doing that they were able to show that the relatively large amplitude BOLD signal peaks \cite{enzo2011} contain substantial information. These findings suggested the possibility that relevant dynamical information can be condensed in \emph{discrete} events.  If that was true, then the possibility to capture space and time was possible, an objective  ultimately achieved in a subsequent report by Tagliazucchi and colleagues \cite{enzo2012} which demonstrated how brain dynamics at resting state can be captured just by the timing and location of such events, i.e., in terms of a spatiotemporal point process.

\subsection{fMRI as a point process}

The application of this novel method allowed, for the first time, to define a theoretical framework in terms of an order and control parameter derived from fMRI data,  where the dynamical regime can be interpreted as  one corresponding to a system close to the critical point of a second order phase transition. The analysis demonstrated that the resting brain spends most of the time near the critical point of such transition and exhibits avalanches of activity ruled by the same dynamical and statistical properties described previously for neuronal events at smaller scales. 

The data in Figure \ref{figpuntos} shows an example of a point process extracted from a BOLD time series. A qualitative comparison with the established method of deconvolving the BOLD signal with the hemodynamics response function suggest that at first order, the point process is equivalent to the peaks of the deconvolucion.

As shown in \cite{enzo2012} the point process can efficiently compress the information needed to reproduce the underlying brain activity in a way comparable with conventional methods such as seed correlation and independent component analysis demonstrated by, for instance, its ability to replicate the right location of each of the RSN.
While the former methods represent averages over the entire data sets, the point process, by construction, compresses and {\it preserves} the temporal information. This potential advantage,  unique of the current approach, may provide additional clues on brain dynamics. 

\begin{figure}
\centerline{
\includegraphics[width=.6\textwidth,clip=true]{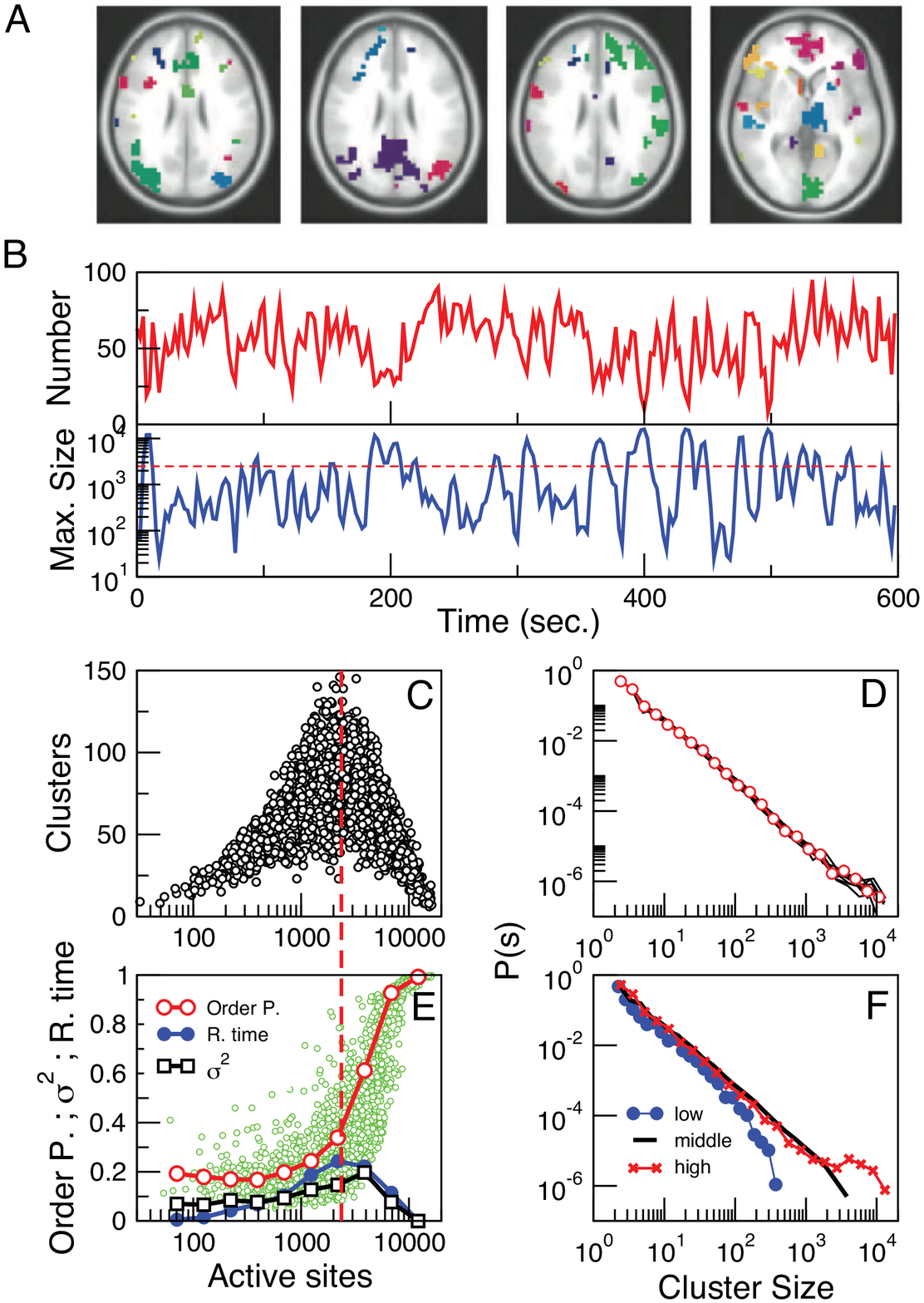}}
 \caption{The level of brain activity continuously fluctuates above and below a  phase transition. (A) Examples  of co-activated clusters of neighbor voxels (clusters are 3D structures, thus seemingly disconnected clusters may have the same color in a 2D slice). (B) Example of the temporal evolution of the number of clusters and its maximum size (in units of voxels) in one individual. (C) Instantaneous relation between the number of clusters vs. the number of active sites (i.e., voxels above the threshold) showing a positive/negative correlation depending whether activity is below/above a critical value ($\sim 2500$ voxels, indicated by the dashed line here and in Panel B). (D) The cluster size distribution follows a power law spanning four orders of magnitude. Individual statistics for each of the ten subjects are plotted with lines and the average with symbols.
 (E) The order parameter, defined here as the (normalized) size of the largest cluster is plotted as a function of the number of active sites (isolated data points denoted by dots, averages plotted with circles joined by lines). The calculation of the residence time density distribution (``R. time'', filled circles) indicates that the brain spends relatively more time near the transition point. Notice that the peak of the R. Time in this panel coincides with the peak of the number of clusters in panel C, as well as  the variance of the order parameter (squares).
(F) The computation of the cluster size distribution calculated for three ranges  of activity (low: 0- 800; middle: 800-5000; and high $>$ 5000) reveals the same scale invariance plotted in panel D for relatively small clusters, but shows changes in the  cutoff for large clusters. Figure redrawn from \cite{enzo2012} 
\label{figenzo2012}}
\end{figure}

This is explored here by compiling the statistics and dynamics of clusters of points both in space and time. Clusters are groups of contiguous voxels with signal above the threshold at a given time, identified by a scanning algorithm in each fMRI volume.  Figure \ref{figenzo2012}A shows examples of clusters (in this case non-consecutive in time) depicted with different colors.  Typically (Fig. \ref{figenzo2012}B top) the number of clusters at any given time varies only an order of magnitude around the mean ($\sim 50$). In contrast, the size of the largest active cluster fluctuates widely, spanning more than four orders of magnitude.

The analysis reveals four novel dynamical aspects of the cluster variability which hardly could have been uncovered with previous methods: 

\begin{itemize}
\item{At any given time, the number of clusters and the total activity (i.e., the number of active voxels) follows a non-linear relation resembling that of percolation \cite{staufer}. At a critical level of global activity ($\sim 2500$ voxels, dashed horizontal line in Fig. \ref{figenzo2012}B, vertical in Fig. \ref{figenzo2012}C) the number of clusters reaches a maximum ($\sim 100-150$), together with its variability.}

\item{The correlation between the number of active sites (an index of total activity)  and the number of clusters reverses above a critical level of activity, a feature already described in other complex systems in which some increasing density competes with limited capacity \cite{Bak,staufer}.}
 
\item{The rate at which the very large clusters (i.e., those above the dashed line in \ref{figenzo2012}B) occurs ($\sim$ one every 30-50 sec) corresponds to the low frequency range at which RSN are typically detected using PICA \cite{beckmann2005}.}
 
\item{The distribution of cluster sizes (Figure \ref{figenzo2012}D) reveals a scale free distribution (whose cut off depends on the activity level, see Panel F).}
\end{itemize}

\subsection{A phase transition}
The four features just described are commonly observed in complex systems undergoing an order-disorder phase transition\cite{Bak,Jensen,chialvo2010}. This scenario was explored in \cite{enzo2012} by defining control and order parameter from the data. To represent the degree of order (i.e. the {\it order} parameter), the size of the {\it largest} cluster (normalized by the number of active sites) in the entire brain was computed and plotted as a function of the number of active points (i.e., the control parameter). This was done for all time steps and plotted in Figure \ref{figenzo2012}E (small circles). 
As a {\it control} parameter  the global level of activity was used  as in other well studied models of order-disorder transitions (the clearest example being percolation \cite{staufer} ).

Several features  in the data reported in \cite{enzo2012} suggest a phase transition: First, there is sharp increase in the average order parameter (empty circles in Fig \ref{figenzo2012}E ), accompanied by an increase of its variability (empty squares). Second, the transition coincides with the peak in the function plotted in Fig. \ref{figenzo2012}C, which accounts for the number of clusters. Finally, the calculation of the relative frequency of the number of  active sites (i.e. the residence time distribution) shows that the brain spends, on the average, more time near the transition than in the two extremes, the highly ordered and the highly disordered states. This supports the earlier conjecture that the brain works near criticality. \cite{Bak,chialvo2010,Expert}. It would be interesting to investigate whether and how this transition diagram changes with arousal states, unhealthy conditions, anesthesia,  etc, as well as to to develop ways to parametrize such changes to be used as objective markers of mind state.
 
\subsection{Variability and criticality} 
It is important to notice that the description in term of a point process allows the observation of activity fluctuations in space and time. In particular note that the  results in (i.e., Fig.\ref{figenzo2012}C and E) show that the resting brain dynamics achieves maximum variability  at a particular level of activation which coincide with criticality. Since  is known that the peak of variability in critical phenomena is found at criticality, it is tempting to speculate that the origin of the brain spontaneous fluctuations can be traced back to a phase transition. This possibility is further strengthened by the fact that the data shows the brain spends most of the time around such transition. 

Thus, overall the results point out to a different class of models which need to  emphasize non-equilibrium self generated variability. The data is orthogonal to most of the current models in which,  without the external noise, the dynamics are stuck in a stable equilibrium state. On the other hand, non-equilibrum systems near criticality do not need the introduction of noise: variability is self-generated by the collective dynamics which spontaneously fluctuate near the critical point.
 
\section{Consequences}

 As discussed in previous sections, critical dynamics implies coherence of activity beyond what is dictated by nearest neighbors connections and correlations longer than that of the neural structure and nontrivial scaling of the fluctuations. These anomalies suggest the need to turn the page on a series of concepts derived from the idea that the brain works as a circuit. While it is not suggested here that such circuits do not exist, fundamentally different conclusions should be extracted from their study.  As a starting point, the following paragraphs will discuss which of the associated notions of connectivity and networks should be revised under the viewpoint of criticality. At the end of the section, an analogy with river beds will be offered to summarize the point.
\subsection{Connectivity vs functional collectivity} The present results suggest that the current interpretation of functional connectivity, an extensive chapter of the brain neuro-imaging literature, should be revised. The three basic concepts in this area are: brain functional connectivity, effective connectivity and structural connectivity \cite{friston,spornsconnectome,horwitz}. The first one ``is defined as the correlations between spatially remote neurophysiological events'' \cite{friston}. Per se, the definition is a statistical one, and it ``is simply a statement about the observed correlations; it does not comment on how these correlations are mediated'' \cite{friston}. The second concept, effective connectivity, is closer to the notion of causation between neuronal connections and ``is defined as the influence one neuronal system exerts over another''. Finally the concept of structural or anatomical connectivity refers to the identifiable physical or structural (synaptic) connections linking neuronal elements.

The problem with the use of these three concepts is that, intentionally or not, they emphasize ``the connections'' between brain regions. This is so,  despite of  cautionary comments emphasizing that ``depending on sensory input, global brain state, or learning, the same structural network can support a wide range of dynamic and cognitive states'' \cite{spornsconnectome}. 

An initial demonstration of the ambiguity in the functional connectivity definition were the results of Fraiman in the Ising model which  explicitly demonstrated\cite{Fraiman2008} the emergence of nontrivial collective states over an otherwise trivial regular lattice (i.e. the Ising's nearest neighbors  structural connectivity). Since is well known that the brain structural connectivity is not a lattice, the replication by the Ising model of many relevant brain networks properties suggested the need to revise our assumptions at the time of interpreting functional connectivity studies.
 
The second blow to the ``connectivity'' framework is given by recent results from Haimovici et al. \cite{haimovici} results. They compared the RSN from human fMRI with numerical results obtained from their network model which is based on the structural connectivity determined earlier by Hagmann et al \cite{hagman}, plus a simple excitable dynamics endowed to each network node. Different dynamics were obtained by changing the excitability of the nodes, but only the results gathered  at criticality compared well with the human fMRI. These striking results indicate that the spatiotemporal brain activity in human RSN represents a collective emergent phenomena exhibited by the underlying structure {\it only at criticality}. By indicating under which specific dynamical conditions the brain structure will produce the empirically observed functional connectivity, Haimovici's results not only re-emphasized that ``the same structural network can support a wide range of dynamic and cognitive states'', but it showed how it can be done. Of course, these modeling results only scratched the surface of the problem, and a theory to deal with dramatic changes in functionality as a function of a global parameters is awaiting.  
 
The third concept in the circuit trio is effective connectivity, which as mentioned above  implies the notion of influence of one neuronal group over another. Implicit to this idea is the notion of causation, which needs to be properly defined to prevent confusion. In this context causation for a given variable boils down to identify which one of all the other co-variables (i.e, degrees of freedom sharing some correlations) predict best its own dynamics. This is done by observing the past states of all the interactions to a given site and estimating which one contribute more to determine the present state of such site. While the idea is always the same, the question of causation can be framed in different ways, by specific modeling, by calculating partial correlations, different variants of Granger causality, transfer entropies, etc.  Independently of the implementation, in systems at criticality, the notion of effective connectivity suffers from severe limitation since emergent phenomena cannot be dissected in the interaction pairs. To illustrate such limitation, it suffices to mention the inability to predict the next avalanche in the sandpile model \cite{Bak87} by computing causation between the nearest neighbors sites.   
  
An important step forward is the work reported recently by Battaglia and colleagues \cite{battaglia} who in the same spirit than in the discussion above begin by stating:
\begin{quote}
The circuits of the brain must perform a daunting amount of functions. But how can ``brain states'' be flexibly
controlled, given that anatomic inter-areal connections can be considered as fixed, on timescales relevant for behavior?
\end{quote}
The authors conjectured, based on dynamical first principles, that even relatively simple circuits (of brain areas) could produce many ``effective circuits'' associated with alternative brain states. In their language, ``effective circuits'' are different effective connectivities arising from circuits with identical structural connectivity. In a proof of principle model, the authors demonstrated convincingly how a small perturbation can change at will from implementing one effective circuit  to another. The effect of the perturbation is, in dynamical terms, a switch to different phase-locking patterns between the local neuronal oscillations. We shall add that, for this switch to be possible, the basins of attraction between patterns need to be close or, in other words, the system parameters need to be tuned to a region near instability. Furthermore they found that ``information follows causality''  which implies that under this conditions brief dynamics perturbations can produce completely different modalities of information routing between brain areas of  a fixed structural network. It is clear that this  is the type of theoretical  framework  needed to tackle the bigger problem of how, at large scale, integration and segregation of information is permanently performed by the brain.  
 
\subsection{Networks, yet another circuit?}

The recent advent of the so called network approach has produced, without any doubt, a tremendous impact across several disciplines. In all cases, accessing the network graph represents the possibility to see the skeleton of the system over which the dynamics evolves, with the consequent simplification of the problem at hand. In this way, the analysis focuses on defining the interaction paths linking the systems degrees of freedoms (i.e., the nodes). The success of this approach in complex systems probably is linked to the universality exhibited by the dynamics of this class of systems.  Universality tells us that, in the same class, in many cases the only relevant information is the interactions, thus in that case a network represents everything needed to understand how they work.

Thus, in the case at hand, the use of network techniques could bring the false hope that knowing the connections between neuronal groups the brain problem will be solved. This illusion will affect even those that are fully aware that this is not possible, because the fascination with the complexity of networks will at least produce an important distraction and delay. The point is that  we could be fulling ourselves in choosing for our particular problem a description of the brain determined by graphs, constructed by nodes, connected by paths, and so on. 

The reflection we suggest is  that, despite changing variables and adopting different names, this new network approach   preserves the same idea that  we consider is (dangerously) rigid for understanding the brain: the concept of a circuit. This notion, introduced as the most accepted neural paradigm for the last century, was adopted by neuroscience from the last engineering revolution (i.e., electronics). Thus, while  is true that action potentials traverse, undoubtedly, and circulate trough paths, the system is not a circuit in the same sense of electronic systems, where nothing unexpected emerges out of the collective interaction of resistors, capacitors and semiconductors.  Thus, if these new ideas will move the field ahead, it will depend heavily on resisting this fascination to prevent the repetition of old paradigms with new names.
 
\subsection{River beds, floods and fuzzy paths}

The question often appears on how the flow of activity during any given behavior could be visualized if the brain operates as a system near criticality.\footnote{When asked, is difficult to resist the temptation to reply by posing another question:   Considering that, according with current ideas, behavior is produced by the activity (action potentials) flowing trough a given circuit,  how the mechanism responsible to switch between one to the other circuit is visualized?}

The answer, in absence of datum, necessarily involves the use of caricatures and analogies. In such hypothetical framework, we imagine a landscape where the activity flows, and to be graphical let think of a river. If the system is near criticality, first and most importantly, such landscape must exclude the presence of deep paths (i.e., no ``Grand Canyon''),  only relatively shallow river beds, some of then with water and some others dry. On the other hand, if the system is ordered the stream will always flow following deep canyons. In this context, let imagine that ``information'' is transmitted by the water, and in that sense it is its flow that ``connects''  regions (whenever at a given time two or more regions are wet simultaneously).  Under relatively constant conditions erosion, due to water flow, will be expected to deepen the river beds. Conversely, changes in the topology of this hypothetical network  can  occur  anytime  that a  sudden increase makes a stream overflow its banks. After that, it will be possible to observe that the water changed course, a condition that  will be stable only until the next flooding. 

Thus, in this loose analogy, the river network structural connectivity (i.e. the relatively deeper river beds) is the less relevant part of the story to predict where information will be shared. The effective connectivity can be created trough the history of the system, and its paths are not even fixed. The moral behind this loose analogy is to direct our attention to the fact that the path's flexibility depends on having a landscape composed by shallow river beds.
  
\section{Summary \& Outlook}

The program reviewed here considers the brain as a dynamical object.  As in other complex systems, the accessible data to be explained are spatiotemporal patterns at various scales.  The question is whether is it possible to explain all these results from a single fundamental principle. And, in case the answer is affirmative, what does this unified explanation of brain activity implies about goal oriented behavior? We  submit that, to a large extent, the problem of the dynamical regime at which the brain operates it is already solved in the context of critical phenomena and phase transitions. Indeed, several fundamental aspects of brain phenomenology have an intriguing counterpart with dynamics seen in other systems when posed at the edge of a second order phase transition.

We have limited our review here to the large scale dynamics of the brain, nevertheless as discussed elsewhere \cite{chialvo2010} similar principles can be demonstrated at other scales.  To be complete, the analysis must incorporate behavioral and cognitive data which will show similar signatures indicative of scale invariance. Finally, and hopefully, overall  these results should give us a handle for a rational classification of healthy and unhealthy mind states.


\begin{thebibliography}{1000}

\bibitem{Bak}Bak P. (1998) {\it How Nature Works, The science of self-organized criticality} {(Copernicus)}

\bibitem{stanley}Stanley HE. (1987) Introduction to phase transitions and critical phenomena.  Oxford Univ. Press.

\bibitem{turing}Turing AM. Computing machines and intelligence.  {\it Mind} {\bf 59}, 236 (1957).



\bibitem{Stepha}Stassinopoulos D, \& Bak P. Democratic reinforcement. A principle for brain function {\it Phys Rev D} {\bf 51} 5033 (1995).

\bibitem{ceccatto}Ceccatto A, Navone H, Waelbroeck  H. Stable criticality in a feedforward neural network {\it Revista Mexicana de F'sica} {\bf 42} 5, 810--825 (1996).

\bibitem{Chialvo99}Chialvo DR \& Bak P. Learning from mistakes.  {\it  Neuroscience } {\bf 90}, 1137 (1999).

\bibitem{Bak2001}Bak P \& Chialvo  DR. Adaptive learning by extremal dynamics and negative feedback.  {\it Phys Rev E  } {\bf 63}, 031912 (2001).

\bibitem{wakelingbak01}Wakeling J \& Bak P. {\it Phys Rev E} {\bf 64}, 051920 (2001).

\bibitem{Bak87} Bak P, Tang C, Wiesenfeld K. Self-organized criticality: An explanation of the 1/f noise. {\it Phys Rev Lett} {\bf 59}, 381 (1987).

\bibitem{Jensen}Jensen HJ. Self-Organized Criticality. Cambridge University Press (1998).

\bibitem{bak1}Bak P. Life laws. {\it Nature} {\bf 391}, 652--653 (1998).

\bibitem{baksneppen}Bak  P \& Sneppen K. Punctuated equilibrium and criticality in a simple model of evolution. {\it Phys Rev Lett} {\bf 71}, 4083--4086 (1993).

\bibitem{chialvo2010} Chialvo DR. Complex emergent neural dynamics. {\it Nature Physics} {\bf 6}, 744--750 (2010).

\bibitem{Bak95}Bak P \&  Paczuski M. Complexity, contingency, and criticality.  {\it Proc Natl Acad Sci U S A  } {\bf 92}, 6689--6696 (1995).

\bibitem{anderson} Anderson P. More is different.  {\it Science } {\bf 4393}, 396 (1972).

\bibitem{Buzsaki}Buzsaki G. Rhythms of the Brain. Oxford University Press (2006).

\bibitem{Linkenkaer}Linkenkaer-Hansen K, Nikouline VV, Palva JM, Ilmoniemi RJ. Long-range temporal correlations and scaling behavior in human brain oscillations.  {\it J Neurosci  } {\bf 21},1370--1377 (2001).

\bibitem{Stam}Stam CJ \& de Bruin EA. Scale-free dynamics of global functional connectivity in the human brain.  {\it Hum Brain Mapp  } {\bf 22}, 97--109 (2004).

\bibitem{Plenz2007}Plenz D \&Thiagarajan TC. The organizing principles of neuronal avalanches: Cell assemblies in the cortex?  {\it Trends Neurosci  } {\bf 30},101--110 (2007).

\bibitem{Bullock}Bullock TH, Mcclune MC,  Enright JT. Are the electroencephalograms mainly rhythmic? Assessment of periodicity in wide-band time series. {\it Neuroscience  } {\bf121}, 233--252  (2003).

\bibitem{Logothetis}Logothetis NK. The neural basis of the blood-oxygen-level-dependent functional magnetic resonance imaging signal.   {\it Philos Trans R Soc Lond B Biol Sci} {\bf 357}, 1003--1037 (2002).

\bibitem{Eckhorn}Eckhorn R. Oscillatory and non-oscillatory synchronizations in the visual cortex and their possible roles in associations of visual features.   {\it Prog Brain Res } {\bf102}, 405--426 (1994).

\bibitem{Miller}Miller KJ, Sorensen LB, Ojemann JG, den Nijs M. Power law scaling in the brain surface electric potential.  {\it PLoS Comput Biol  } {\bf 5}, e1000609. 10.1371/journal.pcbi.1000609 (2009).

\bibitem{Manning}Manning JR, Jacobs J, Fried I,  Kahana MJ. Broadband shifts in local field potential power spectra are correlated with single-neuron spiking in humans.  {\it J Neurosci  } {\bf 29}, 13613--13620 (2009).

\bibitem{Gilden}Gilden DL. Cognitive emissions of 1/f noise.  {\it Psychol Rev } {\bf 108}, 33--56 (2001).

\bibitem{Maylor}Maylor EA, Chater N,  Brown GD. Scale invariance in the retrieval of retrospective and prospective memories.  {\it Psychon Bull Rev  } {\bf 8}, 162--167 (2001).

\bibitem{Ward}Ward LM.  Dynamical Cognitive Science, London: The MIT Press (2002). 

\bibitem{Nakamura}Nakamura T, Kiyono K, Yoshiuchi K, Nakahara R, Struzik ZR, Yamamoto Y.  {\it Phys Rev Lett   } {\bf 99}, 138103 (2007).

\bibitem{Anteneodo}Anteneodo C \& Chialvo DR. Unraveling the fluctuations of animal motor activity.  {\it Chaos} {\bf 19}, 033123 (2009).

\bibitem{Beckers}Beckers R, Deneubourg J-L, Goss S, Pasteels JM. Collective decision making through food recruitment.  {\it Insectes Sociaux} {\bf 37}, 258--267 (1990).

\bibitem{Beekman}Beekman M, Sumpter DJT, Ratnieks FLW. Phase transition between disordered and ordered foraging in PharaohÕs ants.  {\it Proc Natl Acad Sci USA } {\bf 98}, 9703--9706 (2001).

\bibitem{Rauch}Rauch EM, Chialvo DR, Millonas MM. Pattern formation and functionality in swarm models.  {\it Phys Lett A } {\bf 207}, 185--193 (1995).

\bibitem{Nicolis}Nicolis G \& Prigogine I. Self-Organization in nonequilibrium systems: From dissipative structures to order through fluctuations. Wiley, New York (1977).


\bibitem{Takayasu}Takayasu M, Takayasu H, Fukuda K. Dynamic phase transition observed in the internet traffic flow.  {\it Physica A } {\bf 277}, 248--255 (2000).

\bibitem{Lux}Lux T \& Marchesi M. Scaling and criticality in a stochastic multi-agent model of a financial market.  {\it Nature} {\bf 397}, 498--500 (1999).

\bibitem{Malamud}Malamud BD, Morein G, Turcotte DL. Forest fires: An example of self-organized critical behavior.  {\it Science  } {\bf 281},1840--1842 (1998).

 
\bibitem{Peters1}Peters O. \& Neelin D, Critical phenomena in atmospheric precipitation.  {\it Nature Phys  } {\bf 2}, 393--396 (2006).

\bibitem{Peters2}Peters O, Hertlein C, Christensen K. A complexity view of rainfall.  {\it Phys Rev Lett } {\bf 88}, 018701-1 (2002).

\bibitem{Peters3}Peters O \& Christensen K. Rain: relaxations in the sky.  {\it Phys Rev E } {\bf 66}, 036120-1 (2002).


\bibitem{Cavagna}Cavagna A et al, Scale-free correlations in starling flocks. {\it Proc Natl Acad Sci USA} {\bf 107}, 11865--11870 (2010).

\bibitem{eguiluz}Eguiluz VM, Chialvo DR, Cecchi G, Baliki M and Apkarian V. {\it Phys Rev Lett} {\bf 94}, 018102 (2005).  Also as E-print arxiv.org  Cond-mat/0309092.


\bibitem{Fraiman2008}Fraiman D, Balenzuela P, Foss J,  Chialvo DR. Ising-like dynamics in large-scale functional brain networks. {\it Phys Rev E} {\bf 79}, 061922 (2009).


\bibitem{fox2007}Fox MD and Raichle ME.  Spontaneous fluctuations in brain activity observed with functional magnetic resonance imaging. {\it Nat Rev Neurosci} {\bf 8}, 700--711 (2007).

\bibitem{Beckmann-2009}Smith, SM, et al.  Correspondence of the brain's functional architecture during activation and rest. \emph{Proc. Natl. Acad. Sci. U.S.A.} {\bf 106}, 13040--1345 (2009).

\bibitem{beckmann2005}Beckmann CF, De Luca M, Devlin JT, Smith SM.  Investigations into resting-state connectivity using independent component analysis.  {\it Philos Trans R Soc London} {\bf 360}, 1001--1013 (2005).

\bibitem{xiong}Xiong J, Parsons L, Gao J, Fox P. Interregional connectivity to primary motor
cortex revealed using MRI resting state images.  {\it Hum Brain Mapp} {\bf 8}, 151--156 (1999).


\bibitem{cordes}Cordes D et al. Mapping functionally related regions of brain with functional
connectivity MR imaging.  {\it Am J Neuroradiol} {\bf 21}, 1636--1644 (2000).
 

\bibitem{fuku}Fukunaga M et al. Large-amplitude, spatially correlated fluctuations in BOLD
fMRI signals during extended rest and early sleep stages.  {\it Magn Reson Imaging}
{\bf 24}, 979--992 (2006).

\bibitem{vincent} Vincent JL  et al. Intrinsic functional architecture in the anesthetized monkey brain. {\it Nature} {\bf 447}, 83--87 (2007).

\bibitem{Kitz}Kitzbichler MG, Smith ML, Christensen SR, Bullmore E. Broadband criticality of human brain network synchronization. {\it PLoS Comput Biol} {\bf 5}, e1000314 (2009).

\bibitem{Kuramoto} Kuramoto Y. Chemical oscillations, waves and turbulence. Springer, Berlin (1984).

\bibitem{botcharova} Botcharova M, Farmer SF, Berthouze L. A power-law distribution of phase-locking intervals does not imply critical interaction. arXiv:1208.2659 (2012).


\bibitem{Expert} Expert P, Lambiotte R, Chialvo DR, Christensen K, Jensen HJ, Sharp DJ, Turkheimer F. Self-similar correlation function in brain resting-state fMRI.  {\it Journal Royal Soc. Interface} {\bf 8}, 472--479 (2011).


\bibitem{noise}Garret D, Kovacevic N, McIntosh A, Grady C. Blood oxygen level-dependent signal variability is more than just noise. {\it J of Neurosc} {\bf 30}, 4914--4921 (2010).
 
\bibitem{Garrett}Garrett D, Kovacevic N, McIntosh AR, Grady CL. The importance of being variable. {\it Journal of Neuroscience} {\bf 31}, 4496--4503 (2011).


\bibitem{Fraiman2012}Fraiman D \& Chialvo DR. What kind of noise is brain noise: Anomalous scaling behavior of the resting brain activity fluctuations.   \emph{Front Physiol} {\bf 3}, 307 (2012).

\bibitem{enzo2011} Tagliazucchi E, Balenzuela P, Fraiman D, Montoya, P.  Chialvo DR. Spontaneous BOLD event triggered averages for estimating functional connectivity at resting state, {\it Neurosc Lett} {\bf  488}, 158--163 (2011).

\bibitem{enzo2012}Tagliazucchi E, Balenzuela P, Fraiman D, Chialvo DR. Criticality in large-scale brain fMRI dynamics unveiled by a novel point process analysis. {\it Front Physio} {\bf 3}, 15. doi: 10.3389/fphys.2012.00015 (2012).


\bibitem{staufer}Stauffer, D. Aharony A, (1992) Introduction to percolation theory (Taylor \& Francis).

\bibitem{deco} Rolls, E.T \& Deco, G. (2010) The noisy brain. (Oxford University Press, London).

\bibitem{prigogine}Prigogine  I. (1962). Non-Equilibrium Statistical Mechanics. NewYork: Interscience Publishers.

\bibitem{friston}Friston KJ. Functional and effective connectivity in neuroimaging: a synthesis. \emph{Hum Brain Mapp} \textbf{2},  56--78 (1994).

\bibitem{spornsconnectome} Sporns O, Tononi G, Kotter R.
The human connectome: A structural description of the human brain. \emph{PLoS Comput Biol} \textbf{1}, 245--251 (2005).


\bibitem{horwitz}Horwitz B. The elusive concept of brain connectivity. \emph{Neuroimage} \textbf{19},  466--470 (2003).


\bibitem{haimovici}Haimovici A, Tagliazucchi E, Balenzuela P \& Chialvo DR. Brain organization into resting state networks emerges from the connectome at criticality. arXiv:1209.5353 (2012). 
 

\bibitem{hagman}Hagmann P, Cammoun L, Gigandet X, Meuli R, Honey CJ, Wedeen VJ, Sporns O. (2008) \emph{PLoS Biol} \textbf{6}, e159 (2008).

\bibitem{battaglia}Battaglia D, Witt A, Wolf F, Geisel T. Dynamic Effective Connectivity of Inter-Areal Brain Circuits. \emph{PLoS Comput Biol} \textbf{8}, e1002438 (2012)  

\end{thebibliography}
\end{document}